\documentclass{aa}  
\usepackage{graphicx}
\usepackage{txfonts}
\usepackage{color}
\usepackage{wasysym}
\usepackage{natbib}

\begin{document}

\title{The nature and evolution of a-C(:H) nanoparticle substructures \\ and speculations on the origin of the 3-4$\mu$m emission bands}
\titlerunning{The nature and evolution of amorphous hydrocarbon nanoparticles}
        \subtitle{}

\author{A. P. Jones}

   \institute{Institut d'Astrophysique Spatiale, UMR8617, CNRS, Universit\'e Paris-Saclay, B\^at. 121, 91405 Orsay cedex, France\\
             \email{anthony.jones@universite-paris-saclay.fr} }

   \date{Received 5 August 2025 / Accepted 14 September 2025}

  \abstract
   {The nature and evolution of hydrocarbonaceous grains within interstellar and circumstellar media is still far from resolved, perhaps owing to the rather complex nature of their seemingly simple binary atomic compositions.}
   {This work explores the fine details of amorphous hydrocarbon nanoparticle, a-C(:H), composition and the evolution of the inherent sub-structures under extreme conditions, focusing on the characteristic CH$_n$ bands in the $3-4\,\mu$m wavelength region. } 
   {Particular attention is paid to the role of dehydrogenation and its effects on the sp$^3$ and sp$^2$ hybridisations, leading to an extensive conjugated domain functionalisation of the contiguous structural network within a-C(:H) nanoparticles.}
   {Qualitatively this approach is able to explain the origin and evolution, including the appearance and disappearance, of emission bands observed in the $3-4\,\mu$m wavelength regime without a significant aromatic moiety content within the structures.}
   {A diatomic a-C(:H) phase is likely at the heart of the observed dust evolution in the interstellar medium, and circumstellar and photodissociation regions, as observed at short  wavelengths. It appears that we have some way to go in fully understanding these complex materials. Much laboratory work will be required in order to  elucidate their chemical and structural evolution at nanoparticle sizes under extreme conditions.}
   

   \maketitle

\section{Introduction}
\label{sect_intro}

Carbonaceous materials in their hydrogenated forms are a key component of the dust within the interstellar medium (ISM) and circumstellar media. Numerous models have been built to interpret dust observations \citep[e.g.][]{1992A&A...259..614S,1997ApJ...475..565D,2001ApJ...551..807D,2004ApJS..152..211Z,2011A&A...525A.103C,2013A&A...558A..62J,2014A&A...561A..82S,2017A&A...602A..46J,2021ApJ...917....3D} using an array of carbon forms to successfully match observational data, including: graphite, amorphous carbon, hydrogenated amorphous carbon, polycyclic aromatic hydrocarbons [PAHs], and fullerenes. This material diversity illustrates an inherent degeneracy in the dust modelling process and highlights the fact that the detailed nature and evolution of interstellar (hydro)carbonaceous dust within these media is still far from resolved. The intrinsic difficulty of the problem arises from the rather complex nature of such a seemingly simple, primarily carbon but often binary (C and H), atomic composition material, which encompasses diamond, graphite, fullerenes, graphene, graphene quantum dots, carbon quantum dots, carbon polymeric dots, (hydrogenated) amorphous carbons [a-C(:H)], carbon nanotubes, and PAHs. Further, and in addition to composition, particle size effects are extremely important at nanometre size scales because they impose significant constraints upon their structural make-up \cite[e.g.][]{2012A&A...542A..98J,2025A&A...699A..16J}. 

The ideas and speculations presented here are an extension of the inherent nature of interstellar a-C(:H) nanoparticle dust within the THEMIS framework, The Heterogeneous dust Evolution Model for Interstellar Solids, \citep{2012A&A...540A...1J,2012A&A...540A...2J,2012A&A...542A..98J,2013A&A...558A..62J,2014A&A...565L...9K,2016A&A...588A..44Y,2016A&A...588A..43J,2016RSOS....360221J,2016RSOS....360223J,2016RSOS....360224J,2017A&A...602A..46J,2024A&A...684A..34Y}.  Unlike other widely-used dust  models the THEMIS approach to modelling a-C(:H) interstellar dust \citep{2012A&A...540A...1J,2012A&A...540A...2J,2012A&A...542A..98J,2013A&A...558A..62J,2017A&A...602A..46J,Jones.Ysard.2025} encompasses the view that the dust properties respond to, and reflect, their immediate environment, principally the local radiation field but also the gas density and temperature. The observed spectral variations of carbonaceous dust must therefore be seen in terms of a coupled evolution of its chemical composition, structure, and size and shape distribution. 

This work focuses on the emission bands observed over the $3-4\,\mu$m wavelength range because they generally originate from the smallest particles in the dust size distribution, that is the nanoparticles, that are the most vibrationally and thermally excited and therefore dominate the emitted dust spectrum in this spectral region. These are the particles undergoing the most extreme evolution, especially in photo-dissociation regions (PDRs), and are in all probability heading towards an end-of-the-road structural state before ultimately being completely dismembered through extreme ultraviolet (EUV, $E_{\rm h \nu} \gtrsim 10.2$\,eV, $\lambda \lesssim 121$\,nm) photon-driven bond dissociation, restructuring, and fragmentation. 

This work relies upon an accurate attribution of the CH$_n$ ($n = 0$ to 3) stretching bands of hydrocarbon materials within the $3-4\,\mu$m spectral window. Essential to this are the numerous experimental studies on the nature and evolution of hydrocarbonaceous materials used in the interpretation of astronomical carbon dust observations \citep[e.g.,][]{1986AdPhy..35..317R,1994A&A...281..923J,1998JAP....84.3836R,2004A&A...423..549D,2004A&A...423L..33D,2005A&A...432..895D,2008A&A...490..665P,2008ApJ...682L.101M,2012A&A...537A..27G,2012A&A...548A..40C,2025MNRAS.538..266G}. 

This paper is structured as follows: 
Section \ref{sect_intro_aCH} presents some analogues of a-C(:H) sub-structures, 
Section \ref{sect_evolution} details their likely evolution, 
Section \ref{sect_3p3_band} speculates on the origins of the $3.3\,\mu$m band, 
Section \ref{sect_CC_network} summarises the CC network evolution and makes some experimental suggestions, and
Section \ref{sect_astro} discusses the astrophysical consequences.

\section{a-C(:H) nanoparticle sub-structure analogues}
\label{sect_intro_aCH}

\begin{table*}[t]
\caption{The IR band assignments and central wavelengths ($\mu$m) for the  CH stretching modes in the conjugated cyclic molecules shown in Fig~\ref{fig_spectra}.} 
\label{table_cyclos}
\begin{center}
\begin{tabular}{lcccccc}
\hline
\hline
                 &        &      &                             &               &               &          \\[-0.35cm]
 \hspace{1.2cm} origin $\rightarrow$        & sp$^2$  & sp$^2$ &  sp$^3$  & sp$^2$ + sp$^3$                     &   \multicolumn{2}{c}{sp$^2$ + sp$^3$}  \\[0.05cm]
   & aromatic           &  cis                         & paired $2^\circ$ CH$_2$        & $2^\circ$ CH$_2$                    & \multicolumn{2}{c}{paired}   \\[0.05cm]
   molecule    &      CH               &  $-$HC$=$CH$-^{(a)}$  & $-$CH$_2-$CH$_2$$-^{(b)}$          & $=$CH$-$CH$_2$$-$HC$=^{(c)}$  & \multicolumn{2}{c}{$=$CH$-$CH$_2$$-^{(d)}$}  \\[0.05cm]
\hline
                  &      &      &       &     &       \\[-0.3cm]  
 1,3-cyclohexadiene   &            &  3.27 &   3.40  &           &  3.47  & 3.53     \\
 1,4-cyclohexadiene   &            &  3.29 &            &  3.45 &  3.47  &  3.51      \\
 cyclooctatetraene     &             &  3.31 &            &              &             &          \\
 naphthalene              &  3.26 &            &            &              &             &         \\
 anthracene                &  3.28 &            &            &              &             &         \\
 phenanthrene            &  3.32 &            &            &              &             &         \\
 pyrene                       &  3.28 &            &            &              &             &          \\
         &       &      &       &        &      &         \\[-0.3cm]
\hline
        &       &      &       &        &      &         \\[-0.3cm]
 average                      & 3.28 &  3.29  &  3.40  &   3.45  &  3.47  &  3.52                  \\
 $\approx$ dispersion  &   $\pm0.02$  &  $\pm0.02$ &                    &                    &       &   $\pm0.01$                \\[-0.05cm]
        &       &      &       &        &      &         \\[-0.3cm]
\hline
       &       &      &       &        &      &         \\[-0.3cm]
 IS, PDR \& CS bands                        &   \multicolumn{2}{c}{3.29}            & 3.40               & \multicolumn{2}{c}{\hspace{1.0cm} 3.46}            &  {3.51}     \\
 $\approx$ dispersion                    &   \multicolumn{2}{c}{$\pm0.01$}  & $\pm0.01$     & \multicolumn{2}{c}{\hspace{1.0cm} $\pm0.02$}  & $\pm0.02$    \\[0.05cm]
\hline
\hline
\end{tabular} 
\tablefoot{Found in: 
\tablefoottext{a}{in all three cycloalkenes,} 
\tablefoottext{b}{only in 1,3-cyclohexadiene,} 
\tablefoottext{c}{only in 1,4-cyclohexadiene, and} 
\tablefoottext{d}{in 1,3- and 1,4-cyclohexadiene,}
See the text for a justification of the band assignments, where aromatic and olefinic (sp$^2$), and aliphatic (sp$^3$) origins are indicated. All the tabulated bands are comparatively strong. The positions of the observed interstellar emission bands are also indicated; the larger uncertainties for the 3.46 and $3.51\mu$m bands reflect the difficulty in determining the band centre for these weaker and broader bands.} 
\end{center} 
\end{table*}

\begin{figure*}[t]
\centering\includegraphics[width=18.0cm]{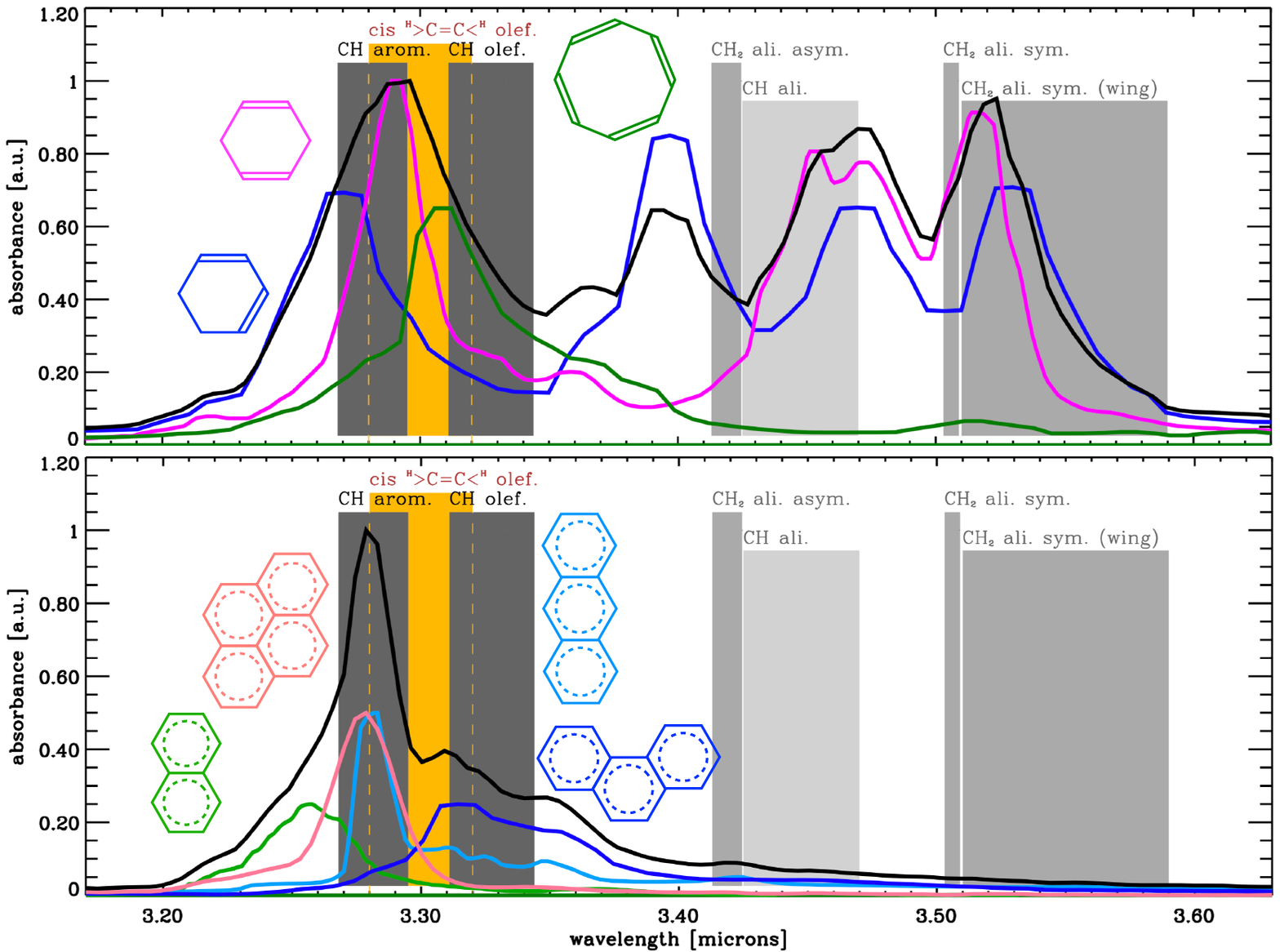}
\caption{The spectra of a selected set of conjugated cyclic and aromatic molecules. Upper panel, conjugated, non-planar molecules 1,3-cyclohexadiene (blue), 1,4-cyclohexadiene (violet), and cyclooctatetraene (green) (see also Fig. \ref{fig_cyclic_structures}). Lower panel, planar, aromatic molecules naphthalene (green), pyrene (rose), antthracene (cobalt), and phenanthrene (blue). The black lines are the normalised spectra of the blended sub-components in the relative fractions $\frac{1}{2}, \frac{1}{4}, \frac{1}{4}$ (upper) and $\frac{1}{6}, \frac{1}{3}, \frac{1}{6}, \frac{1}{3}$ (lower), respectively. The yellow and grey boxes give the usual ranges for the functional groups: aromatic and olefinic CH, olefinic cis $^{\rm H}$\hspace{-0.1cm}$>$C$=$C$<^{\rm H}$, the aliphatic CH$_2$ symmetric and antisymmetric modes, and tertiary ($3^\circ$) aliphatic CH.}
\label{fig_spectra}
\end{figure*}

\begin{figure}[t]
\centering\includegraphics[width=9.0cm]{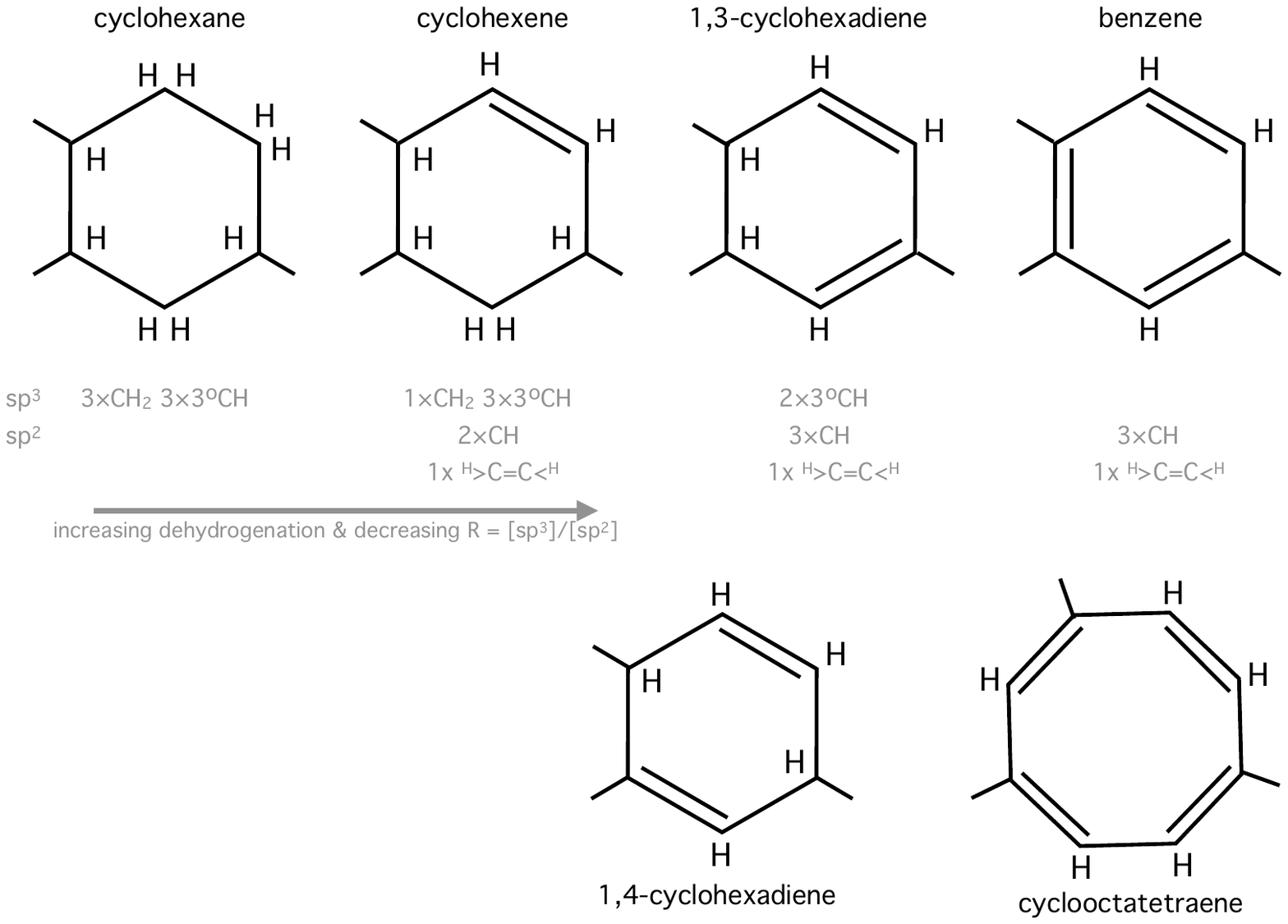}
\caption{A dehydrogenation scenario for the cyclic structures within a-C(:H), starting with a cyclohexane-like ring and evolving through cyclohexene-like, cyclohexadiene-like and ending with an aromatic benzene-like ring. The H atoms inside (outside) the rings are tertiary, $3^\circ$ sp$^3$ (sp$^2$) CH groups and the paired H atoms are sp$^3$ CH$_2$ groups.}
\label{fig_cycles_dehydro}
\end{figure}

\begin{figure}[h]
\centering\includegraphics[width=9.0cm]{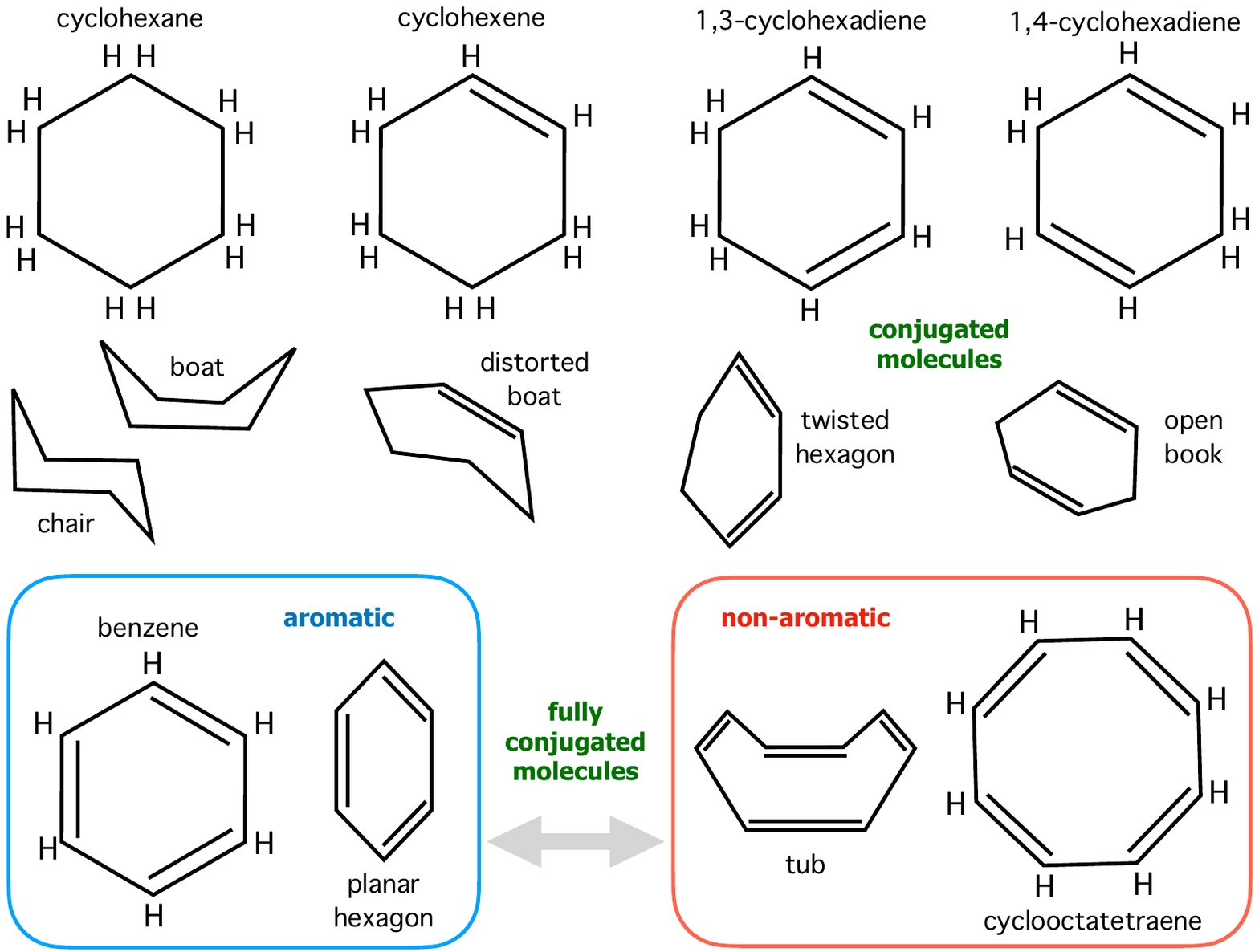}
\caption{The structure of cyclic molecules containing C=C bonds. Benzene is the only planar, aromatic structure and all other molecules have only sp$^3$ $>$CH$_2$ and/or sp$^2$ cis olefinic $^{\rm H}$\hspace{-0.1cm}$>$C$=$C$<^{\rm H}$ groups.}
\label{fig_cyclic_structures}
\end{figure}

The data used in this analysis are summarised in Table \ref{table_cyclos} and the corresponding $3-4\,\mu$m spectra are presented in Fig. \ref{fig_spectra}\footnote{All molecular infrared data are taken from the NIST database https://webbook.nist.gov/chemistry/name-ser/}. In Table \ref{table_cyclos} the bands have been grouped and attributed according to the associations inferred from the spectra of a selected group of conjugated olefinic-aliphatic (1,3-cyclohexadiene, C$_6$H$_8$,  1,4-cyclohexadiene, C$_6$H$_8$, and cyclooctatetraene, C$_8$H$_8$)\footnote{The selected cyclic  olefinic-aliphatic species exhibit only =CH$-$ and $-$CH$_2$$-$ functional groups and no methyl groups ($-$CH3) in order to simplify the analysis. Methyl groups are in any case probably not a significant component of interstellar hydrocarbon dust.} and aromatic molecules (naphthalene, C$_{10}$H$_{8}$,  antthracene, C$_{14}$H$_{10}$, phenanthrene, C$_{14}$H$_{10}$, and pyrene, C$_{16}$H$_{10}$). These cyclic molecules were chosen to be representative of the constituent olefinic-aliphatic sub-structures and the most abundant aromatic moieties within a-C(:H) nanoparticles \cite[e.g.,][]{2012A&A...542A..98J,Jones.Ysard.2025}. It is of note that none of these molecules would be long-lived in the diffuse ISM, PDRs, or H\,{\small II} regions because their lifetimes in these regions are significantly shorter that that of the parent nanoparticles against photo-destruction within those regions 
\citep[e.g.][]{1996A&A...305..602A,1996A&A...305..616A,1997A&A...323..968A,1999ApJ...512..500J,Jones.Ysard.2025}. Their characteristic infrared signatures would therefore only be observable if they were to form part of a larger and more stable molecule or nanoparticle. 

The CH absorption bands associated with these seven selected molecules can be grouped by their central wavelengths, to within $\pm 0.02\mu$m. This gives some leeway and allows for the fact that the exact band centres for particular CH$_n$ ($n \leqslant 2$) olefinic, aromatic, or aliphatic configurations depends upon the local environment within a specific molecule or a-C(:H) sub-structural unit. This is perhaps evident in the comparison of the associated band positions, as inferred in this work, and as  measured in the laboratory \cite[e.g.][]{2019A&A...569A.100B}. 

Table~\ref{table_cyclos} and Fig.~\ref{fig_spectra} indicate that the spectra of these molecular systems exhibit broad groups of bands (in terms of their intensity and shape characteristics), for the aromatics these are 
\vspace{-0.2cm} 
\begin{itemize}
\item $3.26-3.32\mu$m         \hspace*{0.3cm} (strong, narrow), and 
\item $\sim 3.35\mu$m         \hspace*{0.95cm} (weak shoulder),
\end{itemize}
\vspace{-0.2cm} 
and for the aliphatics and olefinics
\vspace{-0.2cm} 
\begin{itemize}
\item $3.28-3.32\mu$m        \hspace*{0.3cm} (strong, broad), 
\item $\sim 3.36\mu$m         \hspace*{0.95cm} (weak shoulder),  
\item $3.39-3.41\mu$m        \hspace*{0.3cm} (strong, broad),  
\item $\sim 3.45\mu$m        \hspace*{0.95cm} (weak, narrow)
\item $\sim 3.47\mu$m        \hspace*{0.95cm} (strong, broad), and
\item $3.51-3.53\mu$m       \hspace*{0.3cm} (strong, broad).  
\end{itemize}
\vspace{-0.2cm} 
It is notable that the dispersions in the band centres are at most $\pm 0.02$, which gives some confidence that the so-grouped bands in this suite of molecules can indeed be assigned to particular CH$_n$ group configurations. It is also notable that the positions of these bands, if not their intensities, bear a remarkable resemblance to the positions of the observed IR emission bands (also shown in Table~\ref{table_cyclos}). The attributions of certain of these IR CH stretching bands are widely accepted \citep[e.g.][]{2019A&A...569A.100B}:   
\vspace{-0.2cm} 
\begin{itemize}
\item $3.29\mu$m aromatic CH \hspace*{1.2cm} ($\sim 3.25\,\mu$m, benzene) 
\item $3.42\mu$m asymmetric CH$_2$  \hspace*{0.65cm} ($3.42\pm 0.01\,\mu$m) 
\item $3.47\mu$m tertiary aliphatic CH \hspace*{0.12cm} ($3.43-3.47\,\mu$m)  
\item $3.50\mu$m symmetric CH$_2$  \hspace*{0.8cm} ($3.50\pm 0.01\,\mu$m).    
\end{itemize}
\vspace{-0.2cm} 
However, and as presented in this study, some of these attributions may be somewhat ambiguous and are not necessarily unique. 

The aromatic systems considered in this study are small because they best represent the most likely structures to be found as the aromatic domains within a-C(:H) materials at nanometre particle size scales \citep[e.g.][]{2012A&A...540A...1J}. We thus consider all of the expected two (naphthalene) and three (anthrracene, phenanthrene) ring, and the most compact four ring (pyrene), structures. Benzene, not studied here, shows three narrow bands in the $\simeq 3.25\mu$m region due to aromatic CH bands. A broader band at these same wavelengths is a characteristic of the spectra of co-added fully aromatic molecules \citep{2013ApJS..205....8S}, which appears to be inconsistent with the interstellar PAH hypothesis. However, the small $2-4$ ring aromatics studied here have bands in the range $3.26-3.32\mu$m region and a blend of their spectra (see Fig. \ref{fig_spectra}) is in better agreement with observations of the $3.3\,\mu$m band. Further, the blend of the three conjugated molecules shown in the upper panel of Fig. \ref{fig_spectra} also gives a band seemingly in good agreement with the observed position of the $3.3\mu$m band. Thus, and despite the very widely-held view, it appears that aromatic molecules may not be the only possible interpretation of the observed $3.3\mu$m IR emission band, which is seemingly composed of two sub-bands \citep{2003MNRAS.346L...1S} of possible sp$^2$ olefinic and/or aromatic origin \citep{2017ApJ...845..123S,2019A&A...569A.100B}.

\section{a-C(:H) nanoparticle sub-structure evolution}
\label{sect_evolution}

In the diffuse ISM all dust species are subject to prolonged irradiation by UV and EUV photons and also to ion irradiation in interstellar shock waves and through cosmic rays. Siliceous materials are rather resistant to irradiation but carbonaceous materials, particularly those containing a significant hydrogen atomic fraction, that is aliphatic-rich a-C:H, are susceptible to energetic photons and ions. This photolysis or photo-processing (or radiolysis in ion irradiation) acts directly through dissociation, or indirectly through heating and thermal processing, leading to  changes in the chemical, structural and physical properties of hydrocarbon materials and thence to changes in their optical and spectral properties. An experimental and theoretical understanding of photo-processing and ion irradiation effects on amorphous hydrocarbon materials is key to advancing our understanding of how these materials evolve under ISM conditions \citep[e.g.,][]{2008A&A...490..665P,2010A&A...519A..39G,2011A&A...529A.146G,2012A&A...537A..27G,2012A&A...548A..40C,2012A&A...540A...1J,2012A&A...540A...2J,2012A&A...542A..98J,2014A&A...569A.119A,2015A&A...577A..16P,2015A&A...584A.123A}.

The primary processes operating during the photolysis of a-C(:H) solids are the removal of H atoms (dehydrogenation)  and the associated transformation of saturated, aliphatic, single sp$^3$ C--C bonds into unsaturated, olefinic double C$=$C and aromatic C$\simeq$C sp$^2$ bonds.  
In order grasp the essential and fundamental elements of how exactly carbonaceous materials evolve chemically and structurally we go back to the basics and consider the chemical building blocks of a-C:H solids. It is clear that (poly-)cyclic species, such as those found in asphaltenes, are an important component in their structures \citep[e.g.,][]{1986AdPhy..35..317R,1988PMagL..57..143R,2012A&A...540A...1J,2012A&A...540A...2J,2012A&A...542A..98J,2013MNRAS.429.3025C}. Considering isolated six-fold ring systems, starting with cyclohexane-type structures (see Fig~\ref{fig_cycles_dehydro}), we can understand how these simple systems evolve as a result of dehydrogenation processes. A likely sequence for this radiative processing is shown in Fig~\ref{fig_cycles_dehydro}, where step-wise dehydrogenation transforms a single six-fold aliphatic cycle into an aromatic benzene-type structure with the removal of a pair of hydrogen atoms in each step. Experimental evidence points to the removal of pairs of atoms as molecular hydrogen under the effects of photolysis, ion irradiation and thermal annealing \citep{1984JAP....55..764S,1989JAP....66.3248A,1996MCP...46...198M,2014A&A...569A.119A}. 

Given that IR spectra are the well-determined, diagnostic chemical fingerprints for the CH$_n$ ($n = 1-3$) vibrational modes in hydrocarbons we can compare their spectra in the $3-4\,\mu$m wavelength region in order to elucidate the characteristic signatures of the CH$_n$ bonding structures to be found within a-C(:H) nano-grains in the ISM. To do this we consider the attribution of the CH stretching modes using a recent graphically-illustrated compilation \cite[][Fig. 1]{2019A&A...569A.100B} and also the spectra of the selected molecules in Fig. \ref{fig_spectra}. Table \ref{table_cyclo_struc} summarises this sequence and shows how the number and types of constituent functional groups evolve with the dehydrogenation steps. From top to bottom and left to right this table indicates the disappearance of aliphatic CH$_2$ groups and appearance olefinic CH configurations, leading to a progressive replacement of the aliphatic CH$_2$ (3.41--3.43, 3.50--3.51$\mu$m) and CH (3.43--3.47$\mu$m) modes with olefinic modes (3.31--3.34$\mu$m).

\begin{table*}[t]
\caption{Analysis of the CH bonding characteristics of isolated hexa-cyclic molecules and the octa-cyclic molecule cyclooctatetraene.} 
\label{table_cyclo_struc}
\centering
\begin{tabular}{lcccccc}
\hline
\hline
                 &        &      &                             &               &               &               \\[-0.25cm]
&  \multicolumn{6}{c}{No. of  distinct groups per molecule ( per a-C(:H) sub-structure )} \\
                                   &                        &  adjacent                     &   3$^\circ$   &  aliphatic                  &                            &                     \\ 
                                    &  aliphatic        & alphatic                      &  aliphatic     &  olefinic                     &  olefinic              &  olefinic  cis    \\
                                    & $>$CH$_2$    &   $-$CH$_2-$CH$_2-$  &  $>$CH$-$  &    $=$CH$-$CH$_2-$    & =C$<^{\rm H}$   &    $^{\rm H}\hspace{-0.1cm}>$C=C$<^{\rm H}$             \\
\hline
        &       &      &       &        &      &     \\[-0.25cm]
 IR band(s)                 &  3.41$-$3.43$_{\rm asym.}$ & 3.41$-$3.43$_{\rm asym.}$  & 3.43$-$3.47 & 3.47, 3.53  & 3.31$-$3.34 &  3.31$-$3.34  \\[0.05cm]
  \ \ \ [$\mu$m]            &  3.50$-$3.51$_{\rm sym.}$ & 3.50$-$3.51$_{\rm sym.}$  &                   &                  &                      &                                  \\[0.05cm]
                                   &               &          &                   &             &                &               \\[-0.3cm]
\hline
                  &      &      &       &     &   \\[-0.25cm]  
 cyclohexane              &   6 ( 3 )           &  0 ( 3 )                           &  0 ( 3 )      &                                       &                        &                      \\
 cyclohexene              &   4 ( 1 )           &  0 ( 1 )                           &  0 ( 3 )      &  2  ( 2 )                           &   2 ( 2 )           &    1 ( 1 )            \\
 1,3-cyclohexadiene   &   2 ( 0 )           &                                      &  0 ( 2 )      &  2  ( 2 )                           &   4  ( 3 )          &    2 ( 1 )            \\
 1,4-cyclohexadiene   &   2 ( 0 )           &                                      &                  &  4  ( 0 )                           &   4  ( 3 )           &   2 ( 1 )              \\
 cyclooctatetraene      &                       &                                      &                  &                                       &   8 ( 5 )            &   4 ( 2 )            \\[0.05cm]
\hline
\hline
\end{tabular}  
\tablefoot{
The numbers in brackets are for the analogous structures within a contiguous a-C(:H) network where three of the CH bonds have been replaced with network-linking C$-$C bonds (see Figs. \ref{fig_cycles_dehydro} and \ref{fig_cyclic_structures}). The assigned band positions are taken from experimental data \cite[see Table 1 of][]{2019A&A...569A.100B}, except for the 3.47 and 3.53$\mu$m bands attributed to the olefinic-aliphatic group $=$CH$-$CH$_2-$ in this work.}   
\end{table*}

\section{The origin of the $3.3\mu$m emission band}
\label{sect_3p3_band}

In considering the composition, structure and substructures of a-C(:H)  materials it is something of an open question as to the exact form and nature of their constituent polycyclic systems. It is generally considered that in their bulk form these semiconducting materials consist of aliphatic, olefinic and aromatic domains or moieties \citep[e.g.,][]{1986AdPhy..35..317R,1987PhRvB..35.2946R,1988PMagL..57..143R,2004PhilTransRSocLondA..362.2477F}. However, it is not clear that this is also necessarily the case for finite sized particles, particularly when those particles have dimensions of the order of a nanometre or less \citep{2012A&A...540A...1J,2012A&A...540A...2J,2012A&A...542A..98J} as per the a-C(:H) nanoparticles proposed within THEMIS. The details of the structure of particles with, at most only hundreds of atoms, is particularly important because the interstellar PAH hypothesis posits that the observed $3-13\mu$m emission bands are due to PAH molecules with a range of sizes. However, there are several issues with this hypothesis, which appear to have been somewhat put aside over the years. Firstly, perfect, planar PAHs cannot explain the presence of the aliphatic CH$_n$ emission bands in the  $3.4-3.5\mu$m wavelength region, without invoking a significant sp$^3$ carbon component, which must lead to a deformation of the molecular shape and possibly to a loss of planarity. Secondly, it assigns the $3.3\mu$m emission band to aromatic CH bonds, something that is not particularly well supported by laboratory studies, which show that the measured PAH  aromatic CH bands peak over the  $\sim 3.2-3.3\mu$m wavelength region for a wide range of molecules \cite[e.g.][Fig. 3]{2013ApJS..205....8S}. 

Chemistry tells us that there are necessary criteria which determine whether a molecule or a domain that could be aromatic actually is aromatic. Following H\"{u}ckel's rule, in order for a molecule to be aromatic  it must be: cyclic,  planar,  fully conjugated (every atom have $p$ orbitals), and  have $(4n+2)$ $\pi$ electrons (integer $n \geqslant 0$). If a molecular structure does not meet all of these criteria it is unlikely to be aromatic. 

The fact that the interstellar PAH hypothesis needs to invoke sp$^3$ components to explain the observed $3.4\mu$m aliphatic CH$_n$ emission bands is troubling because the introduction of a significant sp$^3$ hybridised carbon component into the ring structure of a PAH will lead to a distortion of its structure and a partial or complete loss of planarity. By the above rules, this would lead to some loss of aromaticity and will affect the characteristic IR band signatures. 

Cyclooctatetraene (see Fig. \ref{fig_cyclic_structures}) is an example of a fully conjugated system,\footnote{A fully conjugated molecule is made up of alternating single and double sp$^2$ carbon-carbon bonds, e.g., --HC=CH--HC=CH--HC=CH--} that contains only cis $^{\rm H}$\hspace{-0.03cm}$>$C$=$C$<^{\rm H}$ groups but no aliphatic methylene (-CH$_2$-) groups, that exhibits a $3.31\mu$m band and is not aromatic because it is non-planar.\footnote{Cyclooctateetraene exists in a tub conformation and is therefore significantly non-planar as shown in Fig. \ref{fig_cyclic_structures}.}  Consequently the CH stretching bands typical of PAHs and peaking in the $\sim 3.27-3.32\,\mu$m wavelength region \cite[e.g.,][]{2019A&A...569A.100B} are complemented by that of conjugated cyclic molecules that also have CH stretching peaks in this same range  (see Fig. \ref{fig_spectra}), for example: cyclooctatetraene (3.31$\mu$m), 1,4-cyclohexadiene (3.30$\mu$m), and 
1,3-cyclohexadiene (3.27$\mu$m). The shapes of these molecules are illustrated in Fig. \ref{fig_cyclic_structures}. Note that none of the selected molecules contain a terminating =CH$_2$ group but only cis $^{\rm H}$\hspace{-0.03cm}$>$C$=$C$<^{\rm H}$ and  -CH$_2$-  (methylene) groups.  Given that they contain only two types of CH functional group (-CH= and  -CH$_2$-) they yield more restricted spectral features in the $3-4\,\mu$m region (see Fig. \ref{fig_spectra}). Thus, and if we only consider the $3-4\mu$m emission bands, it would appear that cyclic, non-aromatic, conjugated sp$^2$ systems could also be a contributor to the strongest of the observed bands in this wavelength region.  

In a similar vein, the sterically-strained and distorted multi-benzenoid, aromatics, in the THEMIS a-C(:H) nanoparticles \citep{2012A&A...542A..98J,2012ApJ...761...35M}, cannot be fully aromatic because they are not perfectly planar and their IR spectra will therefore more closely resemble the highly- or fully-conjugated, polycyclic and linear systems such as the olefinic-aliphatic or  alkene-alkane molecules discussed here. 

Another concern with the interstellar PAH hypothesis is that the observed 3.3 and $11.3\mu$m emission bands are often used as a measure of the charge state of the carriers within the interstellar PAH hypothesis;  with the $3.3\mu$m coming from neutral PAHs and the $11.3/3.3\mu$m band ratio increasing with the ionisation state \cite[e.g.][Section 4.2.1]{2008ARA&A..46..289T}. This same effect can also be attributed to an increase in PAH size, which is also coherent with a decrease in the $6.2/7.7\mu$m band ratio \cite[e.g.,][]{2021MNRAS.504.5287R}. The use of these trends in interpreting the observed emission band ratios rely upon the implicit assumption that the bands  essentially come from single, or at least a restricted size range, of the originating molecules. However, we know from extensive modelling of the interstellar nanoparticle emission that we are in all probability dealing with a broad size distribution of particles exhibiting a range of size-dependent temperatures \cite[e.g.,][]{1992A&A...259..614S,1997ApJ...475..565D,2001ApJ...551..807D,2004ApJS..152..211Z,2011A&A...525A.103C,2013A&A...558A..62J,2014A&A...561A..82S,2017A&A...602A..46J,2021ApJ...917....3D}. Thus, the interpretation of the interstellar carbonaceous dust emission bands has to take into account the size distribution and, hence, the relative band intensities as a function of the temperature distribution across the given size distribution. In other words, the observed 3.3 and $11.3\mu$m emission bands likely do not originate from exactly the same particles, with the longer wavelength bands coming from somewhat larger species. Thus, and dependent upon the underlying model assumptions, this will necessarily skew the arrived at interpretation. As a close look at Fig. 12 of \cite{2013A&A...558A..62J} and Fig. 1 of \cite{2021A&A...649A..18G} show, an increase in the $11.3/3.3\mu$m band ratio  can be explained with THEMIS by an increase in the particle size, for a fixed ISRF, with the ratio increasing substantially (by about a factor of 100) as the minimum particle size increases from 0.4\,nm to 1nm. The changing ratio is always dominated by a decrease in the $3.3\mu$m band intensity. Further, Fig. 11 of \cite{2013A&A...558A..62J} shows that a changing $11.3/3.3\mu$m band ratio could also be attributed to a change in the a-C(:H) nanoparticle composition and structure, particularly for wide band gap a-C(:H) materials ($E_{\rm g} \geqslant 0.5$eV). However, for the materials that are of most interest to us in the diffuse ISM and in PDRs only narrow gap materials ($E_{\rm g}  < 0.2$eV) are of relevance and for them the $11.3/3.3\mu$m band ratio varies little for a given size of particle.

\begin{figure*}[t]
\centering\includegraphics[width=18.0cm]{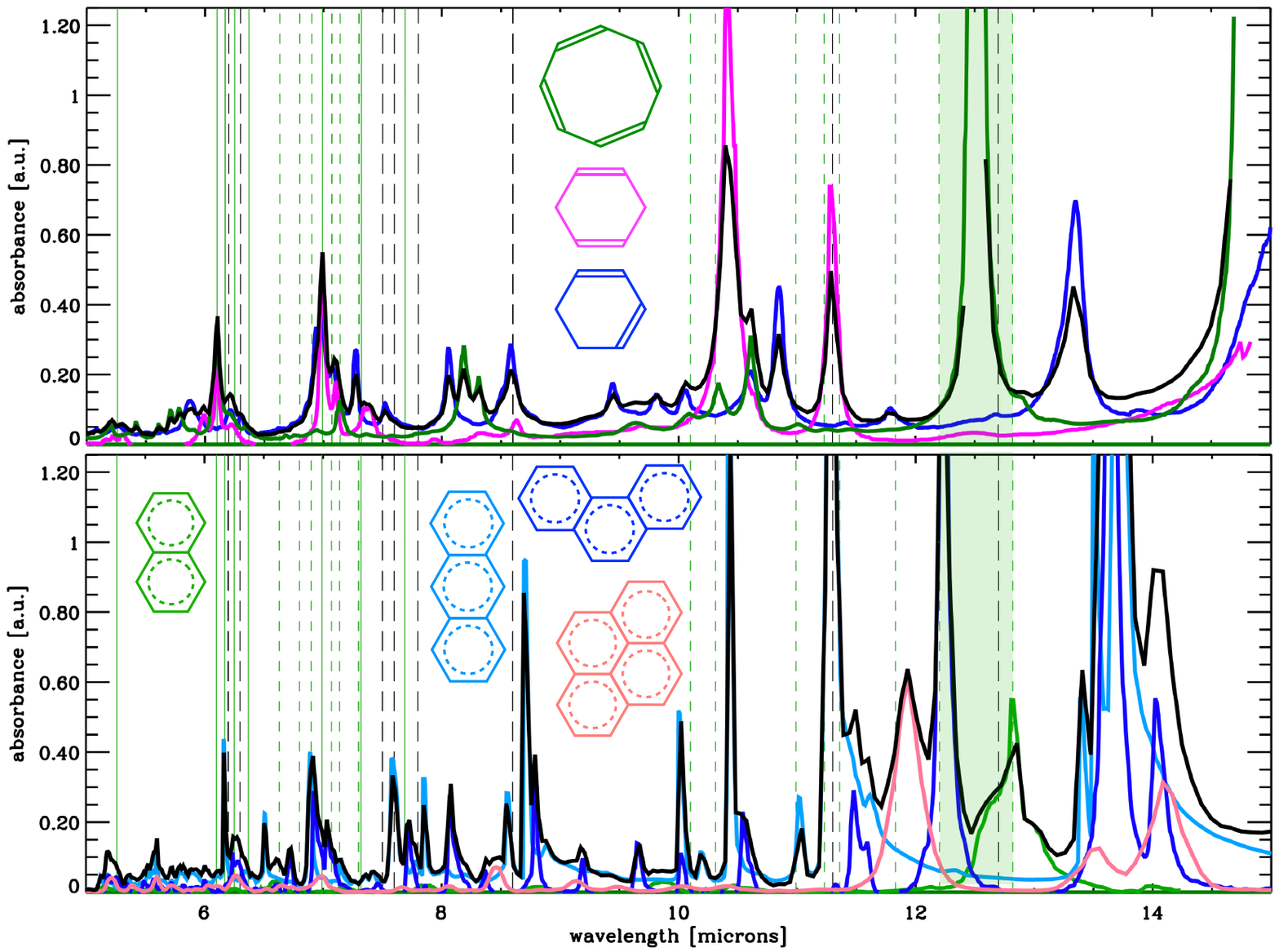}
\caption{The $5-15\mu$m spectrum of the same species, and in the same proportions, as in Fig. \ref{fig_spectra}. The vertical long-dashed black lines indicate the band centres of the interstellar emission bands. The short-dashed (shaded area) and solid green lines show the a-C(:H) modes (range) assigned to CH and CC bonds, respectively \cite[e.g.,][]{2005A&A...432..895D,2008A&A...490..665P,1986AdPhy..35..317R}.}
\label{fig_aromatised}
\end{figure*}

It is evident that a complete study should include a consideration of the longer wavelength emission bands ($\lambda > 4\mu$m) and their relationship with the bands in the $\lambda = 3-4\mu$m region. However, given that we are here most interested in the structure and composition of the smallest interstellar dust particles ($a < 1$nm), that are at the origin of the $3-4\mu$m emission bands, which are undergoing extreme end-of-the-road evolution en route to their eventual destruction, we here concentrate on the bands that most characterise their properties. For this study we therefore do not consider the longer wavelength emission bands, except for a brief discussion in the following section, but focus on the the smallest and hottest nanoparticles which, as in the THEMIS model, do not necessarily have the same structural configurations as their larger counterparts \citep{2012A&A...542A..98J}. The larger particle chemical sub-structures will be more diverse, relaxed and resistant against UV photo-processing. They can therefore retain a wider variety of chemical compositions encompassing less strained cyclic systems including planar aromatic moieties. The bottom line is that it is the smallest interstellar nanoparticles that will, most likely, consist of (sub)structures that exhibit rather extreme forms, particularly in such extreme environments as the diffuse ISM and the Orion Bar and Horse Head PDRs. In these regions they will tend towards the most stable structural configurations possible, likely exhibiting a rather limited range of sub-structures. These are will be highly strained structures that deform what would normally be planar aromatic moieties into distorted, maximally-conjugated, mixed linear and cyclic systems. 

The spectra of the considered hexa-cyclic molecular species 1,4- and 1,3-cyclohexadiene (C$_{6}$H$_{8}$) are shown in Fig.~\ref{fig_spectra}, along with that of the poly-octa-cyclic, fully conjugated molecule cyclooctatetraene  (C$_{8}$H$_{8}$): note that none of the molecules in the upper panel are planar. For comparison the lower panel shows the spectra of a selected set of small (number of aromatic rings, $N_{\rm R} \leqslant 4$), planar, aromatic molecules (naphthalene, anthracene, phenanthrene, and pyrene), all of which are planar. For the conjugated molecules there appears to be a systematic shift in the band position from 3.32 to 3.29 to 3.26$\mu$m, that is with increasing energy as the the rigidity of the structures likely increases (from cyclooctatetraene to 1,4- and then 1,3-cyclohexadiene), which would appear to be coherent with the expected behaviours of IR band positions. Note how the spectrum of the non-planar, non-aromatic, fully conjugated molecule cyclooctatetraene, which contains only -CH= functional groups in cis $^{\rm H}$\hspace{-0.03cm}$>$C$=$C$<^{\rm H}$ and $^{\rm H}$\hspace{-0.03cm}$\geqslant$C$-$C$\leqslant^{\rm H}$ structures, qualitatively resembles that of the aromatic molecules anthracene and phenanthrene, in the sense that the spectrum shows only one predominant band with long wavelength shoulder features. For the blended spectra (black lines), the positions of the 3.3$\mu$m bands occur at 3.29$\mu$m for the conjugated olefinics and at 3.28$\mu$m for the aromatics, which is a small but significant difference. Note that for the conjugated molecules the position of the 1,4-cyclohexadiene band occurs at 3.29$\mu$m, that is, it is almost exactly at the position of the observed interstellar emission band. Also, the band width of the blended spectra appear to be different for these two groups of molecules, being broader for the conjugated olefinic molecule blend where the constituent molecules exhibit intrinsically wider features. Bear in mind that the spectra presented here are for absorbance, whereas the observed interstellar bands are predominantly seen in emission.

\section{Evolution of the CC network bonds and bands}
\label{sect_CC_network}

In their laboratory experiments and IR spectral analysis of more than 50 soot samples \cite{2008A&A...490..665P} find a clear link between the ratio of the 3.3 and $3.4\mu$m CH stretching bands and the position of the CC stretching mode at $6.2\mu$m. The latter mode is characteristic of the nature of the carbon backbone structures within hydrogenated amorphous carbon networks. They find that, as the ratio 3.3/(3.3+3.4) increases, the CC mode shifts from $6.3\mu$m to $6.2\mu$m likely tracing the transformation of carbonaceous dust from aliphatic-rich ($6.3\mu$m) to aromatic-rich ($6.2\mu$m). In the ISM such an effect would be consistent with photo-processing within a given interstellar or circumstellar environment. \cite{2008A&A...490..665P} also find that the most commonly observed IR emission bands, the so-called Class A ($\lambda_{\rm CC} \sim 6.2\mu$m), are most likely composed of more mature, end-of-the-road species, while the less common class C spectra ($\lambda_{\rm CC} \sim6.3\mu$m) appear to be associated with less-processed materials consisting of small aromatic units linked by aliphatic bridges, all embedded within a contiguous network structure.

\subsection{The $5-15\,\mu$m spectrum}
\label{sect_long_wav}

Ideally it would be useful to perform the same kind of exercise as that shown in Fig.~\ref{fig_spectra}, and as discussed above, in order to explore the nature and origin of the $6.2-6.3\mu$m band at a simplified molecular level. However, and because this band is rather a diagnostic of contiguous a-C(:H) network structures, a molecular level comparison would be insufficient. The exact position and form of the $6.2-6.3\mu$m band almost certainly reflects the intimate details of the extended, conjugated sub-structures, that is $-$C$=$C$-$C$=$C$-$C$=$C$-$, to be found within the mixed aliphatic-olefinic and aromatic domains that are key to the contiguous random covalent networks that constitute the extensive family of amorphous (hydro)carbon solids. 

Fig. \ref{fig_aromatised} shows the long wavelength spectrum ($5-15\mu$m) equivalent to Fig. \ref{fig_spectra}. Unfortunately, it is not possible to assign most of these of bands but in order to make comparisons the interstellar emission band centres are indicated by the dashed black lines. Further, and following the work of \cite{2005A&A...432..895D}, \cite{2008A&A...490..665P}, and \cite{1986AdPhy..35..317R} the band positions assigned to the CH and CC bonds in a-C(:H) solids are indicated by the dashed and solid green lines, respectively. 
Not wishing to over-interpret these data it is nevertheless possible to make a few qualitative inferences. It is noteworthy that the aromatics show a somewhat more complex spectrum than the conjugated species. With respect to the interstellar emission bands, features in the 6.2, 8.6 and 11.3$\mu$m regions are evident in the combined spectra of the aromatics and conjugated species, and in the 7.7$\mu$m region are seen in the aromatics, albeit that the band shapes and intensities are very different from those observed in the ISM. With respect to the a-C(:H) assignments, the 6.2$\mu$m region is well populated by CC modes and there is a general correspondence between the CH modes and those seen in the combined spectra. However, many bands in the $6.5-7.5\mu$m region do not seem to correspond to any interstellar bands and there is a strong 7$\mu$m CC mode present in both sets. As discussed in Section \ref{sect_evolution}, the selected set of molecules is over-populated with hydrogen atoms, with respect to a-C(:H) solids, and the CH bands are therefore over-represented in this assignment comparison. Further, the carbon backbone or skeletal modes will be significantly different from the CC modes of the selected molecules because of steric and strain effects due to the incorporation of these kinds of moieties within a 3D a-C(:H) network. This may help to explain the differences noted here between the molecular modes of these molecules and the modes assigned to a-C(:H) random covalent networks.

\subsection{Restructurarion}
\label{sect_restruct}

\begin{figure}[t]
\centering\includegraphics[width=9.0cm]{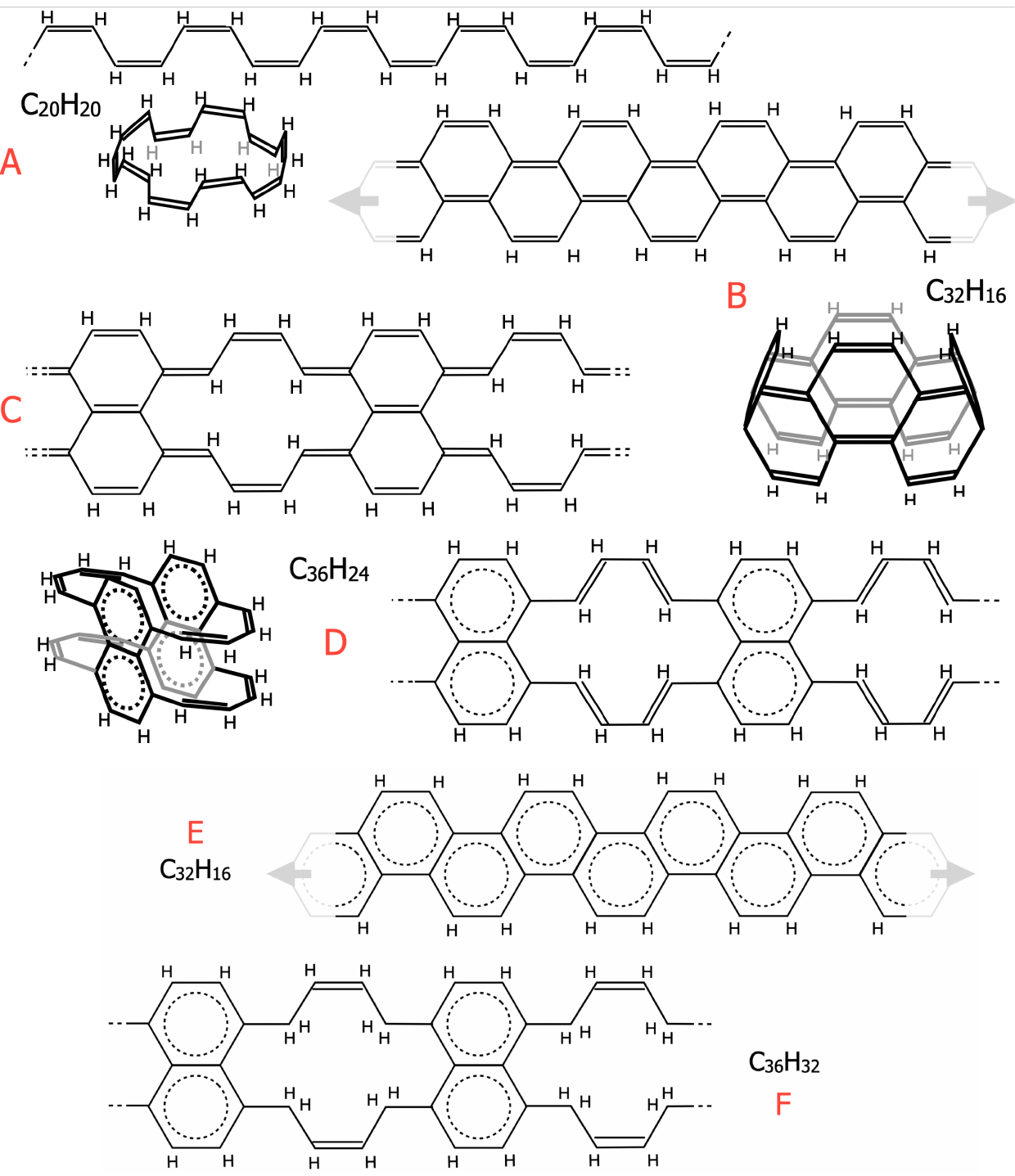}
\vspace{-2.0cm}
\caption{2D examples (thin lines) of fully conjugated 3D systems (thick lines, with the dotted bonds and arrowed/shadowed bonds connected), which exhibit only cis --HC$=$CH--, aromatic CH  and $>$C$=$ functional groups, the latter are sp$^2$ C atoms bonded to three other C atoms.} 
\label{fig_fully_conjugated}
\end{figure}

The structures A to F illustrated in Fig. \ref{fig_fully_conjugated} are 2D representations (thin lines) of idealised 3D closed loop structures (thicker lines), which are maximally conjugated systems exhibiting only cis $^{\rm H}$\hspace{-0.03cm}$>$C$=$C$<^{\rm H}$ and aromatic CH functional groups.\footnote{In their 3D forms all $-$HC$=$CH$-$ groups are cis, even those that appear as trans in their 2D representations (e.g., structure D in Fig. \ref{fig_fully_conjugated}).} Thus, all the olefinic CH bonds in the upper three structures (A, B, and C) should essentially show only a single $3.3\mu$m band. The C structure is shown in a partially aromatic configuration D that will also give a $3.3\mu$m  aromatic CH stretching band, albeit one likely arising from deformed aromatics that are non-planar due to structural strain. 
The transformation of a highly hydrogenated, aliphatic-aromatic form of the C$_{36}$ structure (F, C$_{36}$H$_{32}$) into an olefinic-aromatic form (D, C$_{36}$H$_{24}$)  involves an eight-atom dehydrogenation but its further transformation into an olefinic form (C) involves only an isomerisation of the carbon backbone, as is also required to transform the C$_{36}$H$_{16}$ structure between its fully conjugated (B) and fully aromatic (E) forms. The associated structural transformations towards unsaturated, fully conjugated systems requires C=C bond migration during the rearrangement, something that can occur during UV photocatalysis \cite[e.g.,][]{alkene_isomerisation}. 
It is therefore reasonable to hypothesise that under the effects of interstellar UV photolysis, in the diffuse ISM and in highly excited PDRs and H\,{\small II} regions, that these structures will most likely transform into their more deformable, highly conjugated forms rather than retain a high degree of aromaticity in sterically strained conformations. 

The types of closed loop structures studied here are minimally hydrogenated and will likely be stable against further UV photo-processing dehydrogenation. In reality relatively stable, end-of-the-road structures such as these would probably contain a larger number of atoms than those illustrated here. Additionally, the removal of further hydrogen atoms, and also carbon backbone atoms, would lead to radical structures that have weaker or no CH emission bands and that are therefore not observable in emission at  $3-4\mu$m wavelengths. Although, they may still exist within a given medium and exhibit a $6.2\mu$m band along with any other CC modes at longer wavelengths.

\subsection{Experimental suggestions}
\label{sect_exptl}

This work underlines the need for experimental studies targeted at advancing our understanding of how sub-nanometre hydrocarbon nanoparticles restructure in response to extreme UV (EUV) irradiation ($E_{\rm EUV} \gtrsim 10.2 $eV), which can lead to unexpected structural transformations \citep[e.g.][]{2016RSOS....360223J,alkene_isomerisation}. The inferences made here imply that in highly excited regions the nanoparticles are probably not PAHs but could be fully conjugated, polycyclic, olefinic sp$^2$ systems. Their structural forms might  resemble the ring-shaped segments of carbon nanotubes known as nanobelts \cite[e.g.,][]{doi:10.1021/prechem.3c00083} or take on shell-like forms. 

An experimental protocol designed to investigate the conclusions of this work would have to work with isolated nanoparticles containing less than a few hundred atoms while they are subject to EUV irradiation. Ideally this could be performed in the gas phase using a time of flight apparatus but might, in the first instance, be possible with matrix isolation techniques. As a starting point it should be possible to spectroscopically analyse hydrogenated carbon nanotube fragments (nanobelts) or carbon nanodots  \cite[e.g.][]{2025A&A...699A..16J} at low temperatures before and after EUV irradiation. This could then be coupled with the more difficult task of determining their structural compositions whilst in their ground and excited states, which may be significantly different. 

\section{Astrophysical consequences}
\label{sect_astro}

The $3-4\mu$m spectral region is an important, but not exclusive, diagnostic tool in determining the nature of hydrocarbon dust in the ISM \cite[e.g.,][]{2002ApJS..138...75P}. The dust emission spectra in this wavelength domain are dominated by the so-called aromatic CH band at $3.29\mu$m, which is almost always accompanied by a broad $3.4\mu$m aliphatic band. However, in absorption the $3.29\mu$m band is intrinsically weak compared to the olefinic and aliphatic CH$_n$ modes in this same wavelength region \citep[e.g.][]{2012A&A...542A..98J,2025arXiv250812601P}. 

In this work the term diffuse ISM implies regions where the bands in the $3-4\mu$m window are seen in emission, in contrast to the recent comprehensive study by \cite{2025arXiv250812601P} where the same term is applied to regions where these bands are seen in absorption. These authors suggest that the $3-4\mu$m absorption bands arise from carbonaceous grains with aromatic-rich cores with aliphatic-rich mantles, a grain structure that is consistent with somewhat denser regions of the ISM where cloud- and core-shine are observed \citep{2016A&A...588A..43J,2016A&A...588A..44Y}. 

Given that there seems to be a natural block on the evolution of a-C(:H) solids to materials with band gaps smaller than $\approx 0.2$eV (H atom fraction $X_{\rm H} \simeq 0.05$) during ion or UV irradiation \citep[e.g.,][]{1989JAP....66.3248A,1996MCP...46...198M,2011A&A...528A..56G,2011A&A...529A.146G}, there ought then to be a natural block on their spectral evolution. This could perhaps provide an explanation for the origin and practically ubiquitous observation of the Class A emission band spectra briefly discussed in the preceding section. For these limiting compositions (i.e. $E_{\rm g} \simeq 0.1-0.25$eV) the modelling by \cite{2012A&A...542A..98J} predicts spectra for small particles ($a \lesssim 1$nm) that exhibit an aromatic CH band with satellite aliphatic CH$_{\rm n}$ bands but little evidence for olefinic CH$_n$ bands. Therefore the aromatic CH band should always be associated with aliphatic (but not olefinic) CH$_{\rm n}$ side bands and/or a plateau in the $\approx 3.35-3.55\mu$m region. The lack of olefinic CH bands is because, in the eRCN model \cite{2012A&A...540A...1J,2012A&A...540A...2J,2012A&A...542A..98J}, the sp$^2$ component in low $X_{\rm H}$, sub-nanometre particles is predominantly in small aromatic domains that are linked by aliphatic, rather than olefinic, bridging species. Thus, this modelling predicts that no IR emission band spectrum, arising from hydrocarbon dust, and exhibiting only a $3.29\mu$m aromatic CH band, in the $3-4\mu$m region will be observable in the ISM. Some evidence that seems to support this prediction, and therefore favour an a-C(:H) origin for the IR emission bands, comes from AKARI observations of the IR emission features in M 82, which show no pure aromatic CH band but only aromatic and aliphatic CH bands that always appear together, even in the very energetic super-wind region of M82 \citep{2012A&A...541A..10Y}. In contrast, the PAH and (surface-hydrogenated) graphitic interstellar dust models, implicitly, allow for the possibility of an isolated $3.29\mu$m aromatic CH emission band in the $3-4\mu$m region. Such a $3.29\mu$m emission band has now been observed by JWST in the H\,{\small II} region of the Horshehead Nebula \citep{2025A&A...700A.158M}. As shown in this study, rather than fully aromatic systems, the observation of an isolated $3.29\mu$m band in extreme excitation regions might perhaps be more consistent with an end-of-the-road fully conjugated olefinic, strained nano-structures, containing $\simeq 30-50$ carbon atoms, that are highly dehydrogenated. 

Thus, while large PAHs, possibly with more than 50 carbon atoms, would be stable in the diffuse ISM their aromatic CH stretching bands generally occur in the 3.25$\mu$m region and are therefore probably not a good explanation for the observed 3.3$\mu$m emission band. In contrast, small PAHs with one to four rings and less than 20 carbon atoms give a much better match to the 3.3$\mu$m emission band but, as isolated molecular species, are easily photo-dissociated  in the diffuse ISM. Nevertheless, if they were to exist as sub-structures within larger particles, such as in a-C(:H) nanoparticles, then they would indeed be long-lived enough to contribute to the observed emission bands. However, in this case they are likely to have strained non-planar structures and therefore reduced aromaticity. It is noteworthy that fully conjugated olefinic systems can fit the $3.3\mu$m band equally as well as these one to four ring PAHs.

It is worth noting that the wavelength positions and shapes of the conjugated olefinic -CH$_2$- and -CH= bands, longward of the 3.3$\mu$m band, at 3.40$\pm 0.01$, 3.46$\pm 0.02$, and 3.52$\pm 0.01 \mu$m seen in Fig. \ref{fig_spectra} would appear to provide a good match to the observed Orion  $3-4\mu$m emission spectra \cite[e.g.][]{1997ApJ...474..735S} and to provide a better fit to the more recent James Webb Space Telescope data made using the diffuse ISM THEMIS model \citep{2024A&A...685A..76E}. It is also possible that the wing at 3.56$\pm 0.02 \mu$m, seen in both the laboratory data (see Fig. \ref{fig_spectra}) and the interstellar emission spectra, and as yet unidentified, could be due to contamination from an aldehyde CH functional group (e.g. -C$\leqslant^{\rm H}_{\rm O}$). 

The results of \cite{2003MNRAS.346L...1S} are consistent with a two-component 3.3$\mu$m band composed of a prominant $3.30\mu$m band with a sub-feature at $3.28\mu$m. Both features exhibit a slight shift to shorter wavelength with distance from the central star, HD 44179, coherent with a decreasing temperature of the carriers. The $3.28\mu$m feature and a $3.4\mu$m band are not present in the on-star position but are seen in the offset positions. This points to an evolutionary link between the subordinate 3.28 and $3.4\mu$m features with distance from the star. Assuming emission from nanoparticles with the sub-structure band assignments $\lambda_{\rm \mu m}$[origin]: 3.28[small aromatics], 3.30[conjugated olefinics], and 3.4[aliphatic CH$_2$], then the most stable on-star species is the 3.30$\mu$m band carrier,  perhaps indicating that conjugated olefinic structures may be more stable than aromatic species in highly excited regions. This result seems to underline the intimate association between the aromatic and aliphatic components of carbon nanoparticles \citep[e.g.][]{Jones.Ysard.2025,2012A&A...542A..98J}, point to the stability of conjugated olefinics, and therefore to perhaps somewhat demote interstellar PAHs. We may therefore need to change our current vantage point in order to be able to see interstellar  carbon nanodots for what they really are. 

From this study and in conclusion we may perhaps infer that small, pure, perfect, planar PAHs, with less than $\simeq 30$ C atoms, may not be the dominating nanoparticle dust component that they are often considered to be. It may be that the observable characteristics generally attributed to them (principally the IR emission bands) can be better explained by a distribution of sub-nanometre a-C(:H) particles that, in their final days in extreme environments are actually highly conjugated, polycyclic, olefinic sp$^2$ systems resembling open loop, bowl,  broad ring, or shell structures, which might perhaps be regarded as the hydrogenated segments of carbon nanotubes known as nanobelts \cite[e.g.,][]{doi:10.1021/prechem.3c00083}.

\begin{acknowledgements}
      The author would like to thank the referee for an objective review and encouraging remarks. 
\end{acknowledgements}

\bibliographystyle{aa} 
\bibliography{Ant_bibliography.bib}

\end{document}